\documentclass[]{article}
\usepackage{amsmath}
\usepackage{amsfonts}
\usepackage{amssymb}
\usepackage{graphicx}
\usepackage{amsfonts}
\usepackage[english]{babel}
\usepackage[utf8]{inputenc}
\usepackage{times}
\usepackage[T1]{fontenc}
\usepackage{multirow}

\begin{document}

\title{SO(2,1) Connection in Timelike 3+1 Foliation}

\author{Leonid Perlov, \\
Department of Physics, University of Massachusetts,  Boston, USA\\
leonid.perlov@umb.edu
}

\maketitle

\begin{abstract}
We introduce 3+1 timelike foliation of the four dimensional Lorentz manifold to derive the 3+1 Sen-Ashtekar-Barbero-Immirzi formalism in case of $SO(2,1)$ rotation gauge group, which is possible due to the existence of the $so(2,1)$ algebra isomorphism to $R^3_{2,1}$ algebra with respect to the vector product. We prove that the newly introduced flux and extrinsic curvature variables preserve the symplectic structure of the original variables. We then introduce the modified rotational constraint and succeed to write it as a Gauss constraint of a newly obtained connection. The newly obtained connection is slightly different from the classical 3+1 spacelike Sen-Ashtekar-Barbero-Immirzi connection as it contains in addition the Minkowski metric $\eta_{ij}$ as a coefficient. Our result has a very simple form and clearly shows how $so(2,1)$ connection is different from $so(3)$ one. Also it is the first time that the key-stone fact that makes the whole formalism work in timelike 3+1 case, i.e. $so(2,1) \simeq R^3_{2,1}$ isomorphism and its relation to the $so(2,1)$ connection has been researched.
\end{abstract}
 
\section{Introduction}
 It is known that the Sen-Ashtekar-Barbero-Immirzi connection and flux variables can be introduced only in the 3 dimensional space and do not work for $D > 3$, see $\cite{Thiemann}$. The reason it works in $D = 3$ is due to the existence of the isomorphism between $so(3)$ algebra and  $R^3$ space with the vector product. Such isomorphism does not exist for $D > 3$ and therefore it is impossible to introduce the Sen-Ashtekar-Barbero-Immirzi connection.\\
The question we are trying to find the answer for in this paper is whether the Sen-Ashtekar-Barbero-Immirzi like connection can be introduced for the timelike 3+1 foliation, since there also exists the isomorphism between $so(2,1)$ algebra and the $R^3_{2,1}$ with the corresponding vector product. This gives a hope that the similar formalism might work. Such formalism should include a few steps if done mathematically correctly. We would need to introduce the new $K^i_a$ variables with the values in $so(2,1)$ and the corresponding new $so(2,1)$ rotational constraints $G_i$. We would then need to see if the new $so(2,1)$ variables can be canonically obtained from the original ADM coordinate and momentum variables $(q_{cd}, P^{ab})$ and that the symplectic structure is preserved. If it is preserved we can try to write the new $so(2,1)$ rotational constraint as a Gauss constraint of a new connection by using the algebra $so(2,1) \simeq R^3_{2,1}$ isomorphism. We will have to express the $so(2,1)$ algebra structure coefficients via the $su(2)$ algebra structure coefficients $\epsilon_{ijk}$ and introduce the $so(2,1)$ rotational constraint. Finally we will write the rotational constraint as a Gauss constraint of the newly obtained $so(2,1)$  variables. The  3+1 timelike foliation was previously considered in $\cite{Alexandrov-Kadar}$, $\cite{Alexandrov2}$, \cite{Alexandrov3}, $\cite{Montesinos}$, $\cite{Noui}$, however our result has a much simpler form and clearly shows how $so(2,1)$ connection is different from $so(3)$ one. Also it is the first time that the key-stone fact that makes the whole formalism work in timelike 3+1 case, i.e. $so(2,1) \simeq R^3_{2,1}$ isomorphism and its relation to the $so(2,1)$ connection has been researched. \\[3ex]
The paper is organized as follows. In section $\ref{sec:NewTransformation}$ we introduce the new 3+1 timelike foliation variables. In the next section  $\ref{sec:Isomorphism}$ we discuss the isomorphism of the $so(2,1)$ algebra to $R^3_{2,1}$ vector product algebra. In $\ref{sec:NewConnection}$ we obtain a new connection to be able to write the rotational constraint as a Gauss constraint of that connection.  In section $\ref{sec:SymplecticStructure}$ we prove that the obtained variables are really canonical, i.e. preserving the symplectic structure. $\ref{sec:Discussion}$ concludes the paper. Appendix A ($\ref{sec:AppendixA}$) contains all momentum-momentum Poisson bracket calculation details. Appendix B ($\ref{sec:AppendixB}$) contains all the details of the coordinate-momentum Poisson bracket lengthy calculations.

\section{Timelike 3+1 foliation in $SO(2,1)$ Sen-Ashtekar-Barbero-Immirzi variables}
\label{sec:NewTransformation}
The timelike 3+1 foliation in the regular coordinate-momentum variables requires only one change: $<N, N> = 1$ instead of $<N, N> = -1$ the rest is the same as in the spacelike 3+1 foliation. All the details can be found in \cite{Thiemann}. We are not repeating them here. \\
We start from the timelike foliation in the momentum-coordinate variables and  introduce the $SO(2,1)$ Sen-Ashtekar-Barbero-Immirzi variables in a canonical way. On the four dimensional Lorentz manifold with 3+1 timelike foliation we introduce a bundle space with triads invariant with respect to $SO(2,1)$ rotations and (2,1) metric.
\begin{equation}
\label{q}
q_{ab} = e^i_a e^j_b \eta_{ij}
\end{equation}\\
, where $\eta_{ij}$ is Minkowsky $R^3_{1,2}$ metric $\eta_{ij}=
  \begin{bmatrix}
    1 & 0 & 0  \\
    0 & -1 & 0 \\
    0 &  0 & -1
  \end{bmatrix}
$
\\[2ex]
We introduce the electric flux variable as a weight one density as in $so(3)$ case:
\begin{equation}
\label{E}
E^a_j = {|\det(e)|}e^a_j, \;\;  E^j_a = e^j_a/{|\det(e)|}
\end{equation}
We will use the notation $q = |\det(e^i_a)|^2$\\[2ex]
We then introduce the $K^i_a$ one-form in a little different way than in the spacelike case (notice $\eta_{ij}$):
\begin{equation}
\label{Kab}
K_{ab} := K^i_{(a}e^j_{b)}\eta_{ij}
\end{equation}
satisfying the modified rotational constraint (again notice $\eta_{ij}$):
\begin{equation}
\label{rconstraint}
G_{ab} := K^i_{[a}e^j_{b]}\eta_{ij}  = 0
\end{equation}
, where $\eta_{ij}$ is a Minkowski metric.
By using ($\ref{E}$) we can rewrite it as:
\begin{equation}
\label{rconstraint1}
G_{ab} := K^i_{[a}E^j_{b]}\eta_{ij}  = 0
\end{equation}
or by raising the indices a and b we obtain the form:
\begin{equation}
\label{rconstraint2}
G^{ab} := q^{at}q^{be}K^i_{[t}E^j_{e]}\eta_{ij}  = 0
\end{equation}
Note that both ($\ref{q}$), ($\ref{E}$), ($\ref{Kab}$), ($\ref{rconstraint}$), ($\ref{rconstraint1}$) and ($\ref{rconstraint2}$)  differ from the corresponding $SO(3)$ expressions by the presence of the Minkowski metric $\eta_{ij}$. \\[2ex]
This rotational constraint can be converted into a different but equivalent form by multiplying both sides of the above equality by $e^a_k e^b_m$ and doing summation:
\begin{equation}
\label{jkconstraint}
G_{ab}e^a_k e^b_m = K^i_ae^j_b\eta_{ij}e^a_k e^b_m  - K^i_be^j_a\eta_{ij}e^a_k e^b_m  = K_{aj}e^a_k\delta^j_m - K_{bj}e^ b_m\delta^j_k = 2 K_{a[m}e^a_{k]} = 0
\end{equation}
, which by using ($\ref{E}$) can be rewritten as:
\begin{equation}
\label{jkconstraint2}
G_{jk} = K_{a[j}E^a_{k]}=0
\end{equation}
or by using the antisymmetric tensor as:
\begin{equation}
\label{RotConstraint1}
G_i = \epsilon_{ijk}K_{aj}E^a_{k}=0
\end{equation}
, where in ($\ref{jkconstraint}$) we used: $e^j_be^b_m = \delta^j_m$ , $K^i_a\eta_{ij} = K_{aj}$\\[2ex]
By comparing the two forms of the rotational constraint ($\ref{rconstraint}$) and  ($\ref{jkconstraint2}$) we can see that while the first form contains Minkowski metric $\eta_{ij}$, the second, having been derived from the first does not. This is due to the property $e^j_be^b_m = \delta^j_m$, which follows from $\eta^{ij} = 1/\eta_{ij}$.\\[2ex]
Finally we can rewrite these constraints once again by multiplying each one of $G_i$ by the constants $\eta_{jj}$. We use the sum explicitly to emphasize that there is not summation in $i$ index. Also we use the following identities: $\eta^{ij} = \eta_{ij}$ and $\epsilon_{jkl} = \epsilon^{jkl}$\\[2ex]
\begin{equation}
\label{RotConstraint1}
G_i = \sum_{j,k = 1}^3\eta_{ii}\epsilon_{ijk}K_{aj}E^a_{k}=0
\end{equation}
or we can rewrite it in even better form with all index summation:
\begin{equation}
\label{RotConstraint1}
G_i = \eta_{ij}\epsilon_{jkl}K_{ak}E^a_{l}=0
\end{equation}
Now we have the rotational constraint written by using the $so(2,1)$ algebra structure coefficients. \\
We will use the rotational constraint in the form ($\ref{rconstraint2}$)  when we prove that the transformation to the new variables is canonical (preserving the symplectic structure). The form ($\ref{RotConstraint1}$) of the rotation constraint will be useful when we will be obtaining the new $so(2,1)$ connection.

\section{$so(2,1) \rightarrow R^3_{2,1}$ Isomorphism and Structure coefficients}
\label{sec:Isomorphism}
We are going to show that the constraint $G_{jk}$ can be written as Gauss constraint of $SO(2,1)$ gauge theory.\\
We use the covariant derivative that is compatible with triad: $D_a e^j_b=0$, from which follows $D_a E^j_b=0$.
In case of $E^a_j$ being an $SU(2)$ valued vector density of weight one, the covariant derivative can be written as \cite{Thiemann}:
\begin{equation}
\label{CDerSO3}
D_aE^a_j = [D_aE^a]_j + \Gamma^l_{aj}E^a_l = \partial_aE^a_j + \epsilon_{jkl}\Gamma^k_aE^a_l=0
\end{equation}
This is due to isomorphism between algebra $so(3$) and the Euclidean space $R^3$ with the vector product, as so(3) antisymmetric tensors can be written as 
\begin{equation}
\label{SO(3)Isomorphism}
\Gamma^l_{aj} = \epsilon_{jkl}\Gamma^k_a
\end{equation}
, where $\Gamma^l_a$ is a vector in $R^3$, while  $\epsilon_{jkl}$ are $so(3)$ algebra structure coefficients and we used $\epsilon_{jkl} = \epsilon^{jkl}$ for $so(3)$ case.\\[2ex]
The similar isomorphism exists between $so(2,1)$ algebra and the vectors of the $R^3_{2,1}$ space with respect to its vector product. Since $sl(2,R)$ algebra is isomorphic to $so(2,1)$ algebra by the map: $Y \rightarrow \mbox{ad} \,Y$ we will work with $sl(2,R)$ here. First we express the $sl(2,R)$ algebra structure coefficients via $su(2)$ structure coefficients $\epsilon_{ijk}$. As we could not find such expression in any literature, we derive it here. 
$sl(2,R)$ algebra generators are as follows: \cite{Fomenko}:\\[2ex]
$Y_1=1/2
  \begin{bmatrix}
    0 & 1  \\
    -1 &  0
  \end{bmatrix}, \,
Y_2= 1/2
  \begin{bmatrix}
    1 & 0  \\
    0 &  -1
  \end{bmatrix}, \,
Y_3= 1/2
  \begin{bmatrix}
    0 & 1  \\
    1 &  0
  \end{bmatrix}$\\[2ex]
The commutators of this generators are:\\[2ex]
\begin{equation}
\label{Ycommutators}
 [Y_1, Y_2] = -Y_3,   \;\;   [Y_3, Y_1] = -Y_2,   \;\;  [Y_2, Y_3] = Y_1
 \end{equation}
 or they can be written in a short form by using the algebra structure coefficients as 
 \begin{equation}
 \label{Ystructure}
 [Y_k, Y_l] = \eta_{ii}\epsilon_{ikl}Y_i
 \end{equation}
 or in a better form:
  \begin{equation}
 \label{Ystructure1}
 [Y_k, Y_l] = \eta_{ij}\epsilon_{jkl}Y_i
 \end{equation}
, where $\eta_{ij}$ is Minkowsky $R^3_{1,2}$ metric $\eta_{ij}=
  \begin{bmatrix}
    1 & 0 & 0  \\
    0 & -1 & 0 \\
    0 &  0 & -1
  \end{bmatrix}
$, and we used again $\epsilon_{jkl} = \epsilon^{jkl}, \eta_{ij} = \eta^{ij}$
\\[2ex]
 We calculate the adjoint representation by using the $sl(2,R)$ algebra structure coefficients:\\[2ex]
 $\mbox{ ad } Y_1=
 \begin{bmatrix}
    0 & 0 &0 \\
    0 & 0 & -1 \\
    0 & 1 & 0
  \end{bmatrix},
\mbox{ ad } Y_2=
  \begin{bmatrix}
    0 & 0 &1 \\
    0 & 0 & 0 \\
    1 & 0 & 0
  \end{bmatrix}, 
\mbox{ ad } Y_3= 
  \begin{bmatrix}
    0 & -1 & 0 \\
    -1 &  0 &0 \\
     0 & 0 & 0 
  \end{bmatrix}$\\[2ex]
Or we can write in a general form :
  \begin{equation}
 \mbox{ad} \; Y_k = \eta_{ij}\epsilon_{jkl} 
\end{equation}
\\
These matrices are $so(1,2)$ algebra generators, while $so(1,2)$  is isomorphic to $so(2,1)$.\\[2ex]
Thus the $so(2,1) \simeq R^3_{2,1}$ algebra isomorphism can be written similarly to the $so(3) \simeq R^3$ one  ($\ref{SO(3)Isomorphism}$), by using the map between $so(2,1)$ algebra tensors and $R^3_{2,1}$ vectors connected by the $sl(2,R)$ (or $so(2,1)$) algebra structure coefficients:
\begin{equation}
\label{SO(2,1)Isomorphism}
\Gamma^l_{ai} = \eta_{ij}\epsilon_{jkl}\Gamma^k_a
\end{equation}
or, since $\eta^{ij} = \eta_{ij}$ and $\epsilon_{jkl} = \epsilon^{jkl}$\\[2ex]
\begin{equation}
\Gamma^l_{ai} = \eta_{ij}\epsilon^{jl}_k\Gamma^k_a = \eta_{ij}\epsilon^{jkl}\Gamma^k_a = \eta_{ij}\epsilon_{jkl}\Gamma^k_a
\end{equation}
, where $\Gamma^l_a$ is a vector in $R^3_{2,1}$, while  $\eta_{ij}\epsilon_{jkl}$ are $so(2,1)$ algebra structure coefficients.\\[2ex]

\section{$SO(2,1)$ Timelike foliation Connection}
\label{sec:NewConnection}
By using the $so(2,1) \simeq R^3_{2,1}$  isomorphism ($\ref{SO(2,1)Isomorphism}$)
the covariant derivative ($\ref{CDerSO3}$) then can be rewritten for $so(2,1)$ case as
\begin{equation}
\label{CDerSO21}
D_aE^a_i = [D_aE^a]_i + \Gamma^l_{ai}E^a_l = \partial_aE^a_i + \eta_{ij}\epsilon_{jkl}\Gamma^k_aE^a_l=0
\end{equation}
As in $SO(3)$ case one can easily see that $so(2,1)$ tensors $\Gamma^i_a$ are invariant under Weyl canonical transformation:
\begin{equation}
(K^i_a, E^a_i) \rightarrow ^{(\beta)}K^i_a = \beta K^i_a, \;\; ^{(\beta)}E^a_i = E^a_i/\beta
\end{equation}
The invariance follows from the explicit formula for  $\Gamma^i_a$ expressed via $\Gamma^l_{aj}$ from ($\ref{SO(2,1)Isomorphism}$), while the latter expressed via triads:
\begin{multline}
\Gamma^i_a = \frac{1}{2} \eta^{im}\epsilon^{mjk}e^b_k[e^j_{a,b} - e^j_{b,a} + e^c_je^l_ae^l_{c,b}] 
\\
= \frac{1}{2}\eta^{im}\epsilon^{mjk}E^b_k[E^j_{a,b} - E^j_{b,a} + E^c_jE^l_aE^l_{c,b}]
\\
=\frac{1}{4}\eta^{im}\epsilon^{mjk}E^b_k\left[2E^j_a\frac{(\det(E))_{,b}}{\det(E)} - E^j_b \frac{(\det(E))_{,a}}{\det(E)}\right]
\end{multline}
We can see that $\Gamma^i_a$ is a homogeneous function of the degree zero. \\[2ex]
Therefore $^{(\beta)}\Gamma^j_a= \Gamma^j_a$ and a covariant derivative $D_a$ does not depend on $\beta$ and  $D_a(^{(\beta)}E^a_j) = 0$.\\
The $so(2,1)$ rotational constraint ($\ref{RotConstraint1}$) does not depend on $\beta$ either.
\begin{equation}
G_i = \eta_{ij}\epsilon_{jkl} K_{ka}E^a_l =  \eta_{ij}\epsilon_{jkl} (^{\beta} K_{ka}) (^{\beta}E^a_l)
\end{equation}
 Now we can write the Gauss constraint ($\ref{RotConstraint1}$) as
\begin{multline}
\label{result}
G_i = 0 + \eta_{ij}\epsilon_{jkl}  (^{\beta} K_{ka})  (^{\beta} E^a_l) 
\\
=D_a(^{(\beta)}E^a_i)  + \eta_{ij}\epsilon_{jkl}  (^{\beta} K_{ka})  (^{\beta} E^a_l) 
\\
= \partial_a (^{\beta}E^a_i ) +  \eta_{ij}\epsilon_{jkl}  \Gamma^k_a + \eta_{ij}\epsilon_{jkl}(^{\beta}K_{ka}) (^{\beta}E^a_l)
\\
= \partial_a (^{\beta}E^a_i ) +  \eta_{ij} \epsilon_{jkl}  \left[\Gamma^k_a + (^{\beta}K_{ka}) \right] (^{\beta}E^a_l)
= ^{\beta}\mathcal{D}_a (^{\beta}E^a_i)
\end{multline}
or by introducing notation:
\begin{equation}
\label{so(2,1)connection}
^{\beta}A^l_{ai} = \eta_{ij}\epsilon_{jkl}(\Gamma^k_a + (^{\beta}K_{ka}))
\end{equation}
We can rewrite ($\ref{result}$) as:
\begin{equation}
 ^{\beta}\mathcal{D}_a (^{\beta}E^a_i) =  \partial_a (^{\beta}E^a_i ) +(^{\beta}A^l_{ai})(^{\beta}E^a_l)
\end{equation}
It means that  in $SO(2,1)$ case we can  introduce a new connection, such that the covariant derivative is compatible with the flux $E^a_i$ and the rotational constraints become the Gauss constraint. The $so(2,1)$ connection $(^{\beta}A^k_{ai})$ is different from the $so(3)$ connection by the signs of its components. Interesting that not only the spin connection $\Gamma^k_a$ gets multiplied by the Minkowski metric, but the Barbero-Immirzi part as well. Notice the covariant index k in $K_{ka}$ instead of a contravariant in $so(3)$ case: $K^k_a$, which in $so(3)$ case was easily raised up by Euclidean metric $\delta^{ij}$, however it can not be done same way in $so(2,1)$ case. \\[2ex]
If we consider $\eta_{ij} \epsilon_{jkl}$ as $so(2,1)$ commutator, we can rewrite ($\ref{result}$) as
\begin{equation}
 ^{\beta}\mathcal{D}_a (^{\beta}E^a) =  \partial_aE^a + [A_aE^a]_{so(2,1)}
\end{equation}
similar to the $so(3)$ case with a regular commutator $\epsilon_{jkl}$:
\begin{equation}
 ^{\beta}\mathcal{D}_a (^{\beta}E^a) =  \partial_aE^a + [A_aE^a]_{so(3)}
\end{equation}
The form of the obtained $so(2,1)$ connection ($\ref{so(2,1)connection}$) points to the fact that in 3+1 spacelike $so(3)$ case the metric coefficient should also be present, however, since it is a Euclidean unity matrix, it is of course omitted. Thus the $so(2,1)$ connection differs from $so(3)$ one not only by $\Gamma^k_a$ and $K_{ka}$ taking values in $so(2,1)$ algebra rather than in $so(3)$, but also by the signs of its components due to the presence of the additional Minkowski metric coefficient $\eta_{ij}$. Therefore we can not write the $so(2,1)$ connection in the Sen-Ashtekar-Barbero-Immirzi form:
$A^k_a = \Gamma^k_a + \beta K^k_a$, but only in the full form: $A^l_{ai} = \eta_{ij}\epsilon_{jkl}(\Gamma^k_a + \beta K_{ka})$.
\section{Timelike 3+1 foliation sympectic structure}
\label{sec:SymplecticStructure}
In this chapter we are going to prove that the defined below transformations from the $q_{ab}$ and $P^{cd}$ variables to the new variables $E^a_j$ and $K^j_a$ preserve the the ADM symplectic structure, i.e provided the new variables satisfy:\\
\begin{equation}
\label{Poissontrivial}
\{E^a_j(x), E^b_k(y)\} = \{K^j_a(x), K^k_b(y)\} = 0, \; \{E^a_i(x), K^j_b(y)\} = \frac{k}{2}\delta^a_b\delta^j_i \delta(x,y)
\end{equation}
, where $k= 16\pi G/c^3$ - gravitational coupling constant.\\[2ex]
and provided that all rotational constraints  $G_{ab} = 0$ are satisfied, the variables $q_{ab}$ and $P^{cd}$ satisfy the same Poisson bracket structure as in regular coordinate and momentum coordinates $(q_{ab}, P^{cd})$.\\[2ex]
We define the following transformation from $E^a_j$ and $K^j_a$ variables to $q_{ab}$ and $P_{cd}$ corresponding to 3+1 timelike foliation with the $SO(2,1)$ gauge group. Notice that they are different from the $SO(3)$ ones by the presence of $\eta_{ij}$ in several places. 
\begin{equation}
\label{pandq}
q_{ab} = E^i_aE^j_b\eta_{ij} |\det E^c_i|^{2/D-1}, \;
P^{cd} = 2|\det(E^c_e)|^{-2/(D-1)} ( E^c_kE^{t}_{m}\eta^{mk}K^i_tE^d_i  - E^c_kE^{d}_{m}\eta^{mk}K^i_tE^t_i)
\end{equation}

Below we will calculate the momentum-momentum and momentum-coordinate Poisson brackets. The coordinate-coordinate $\{q_{ab}, q_{cd}\} =0$ is always zero, since $q_{ab}$ contains only electric fluxes $E^a_i$ and the $\{E^a_j(x), E^b_k(y)\} = 0$ as it follows from  ($\ref{Poissontrivial}$) and ($\ref{pandq}$).\\

Let us calculate the Momentum-Momentum Poisson bracket first. We use a new momentum formula ($\ref{pandq}$):

\begin{equation}
P^{ab}(x) = 2|\det(E^c_e)|^{-2/(D-1)}( E^a_kE^{t}_{m}\eta^{mk}K^i_tE^b_i  - E^a_kE^{b}_{m}\eta^{mk}K^i_tE^t_i ) 
\end{equation}

\begin{equation}
P^{cd}(x) = 2|\det(E^c_e)|^{-2/(D-1)} ( E^c_kE^{t}_{m}\eta^{mk}K^i_tE^d_i  - E^c_kE^{d}_{m}\eta^{mk}K^i_tE^t_i)
\end{equation}
\\

In the following we use the notation: $ q:= |\det(E^c_e)|^{2/(D-1)} = |\det e |^2$\\

\begin{multline}
\label{momentumbracket4}
\{P^{ab}(x), P^{cd}(y)\} = \{\frac{2}{q}(E^a_{k_1}E^{t_1}_{p_1}\eta^{p_1k_1}K^{i_1}_{t_1}E^b_{i_1} - E^a_{k_2}E^{b}_{p_2}\eta^{p_2k_2}K^{i_2}_{t_2}E^{t_2}_{i_2}) , \\
 \frac{2}{q}(E^c_{m_1}E^{e_1}_{p_3}\eta^{p_3m_1}K^{j_1}_{e_1}E^d_{j_1} - E^c_{m_2}E^{d}_{p_4}\eta^{p_4m_2}K^{j_2}_{e_2}E^{e_2}_{j_2})
\end{multline}
We provide the detailed calculations of this Poisson bracket in the Appendix A ($\ref{sec:AppendixA}$). The result of the lengthy calculations is as follows:
\begin{equation}
\label{momentumbracket11}
 \{P^{ab}(x), P^{cd}(y)\} = 2kqq^{ac}G^{bd}
\end{equation}
, where \\[2ex]
$G^{bd} =q^{bt}q^{dp}K^{j}_{t}E^{i}_{p}\eta_{ij} - q^{dt}q^{bp}K^{j}_{t}E^{i}_{p}\eta_{ij}= K^{bj}E^{di}\eta_{ij} - K^{dj}E^{bi}\eta_{ij} =2K^{j[b}E^{d]i}\eta_{ij}$\\[2ex]
are the $so(2,1)$ rotational constraints ($\ref{rconstraint2}$)  .\\[2ex]
When the rotational constraint in ($\ref{momentumbracket11}$) is zero, the Poisson bracket is zero. So the momentum-momentum Poisson bracket remains the same as in the original variables.\\[2ex]
Now we turn to the Coordinate-Momentum bracket:
\begin{equation}
P^{ab}(x) = 2|\det(E^c_e)|^{-2/(D-1)}( E^a_{k}E^{t}_{m}\eta^{mk}K^{i}_{t}E^b_{i} - E^a_{k}E^{b}_{m}\eta^{mk}K^{i}_{t}E^{t}_{i} ) 
\end{equation}
 \begin{equation}
 q_{cd}(y) = E^{j}_cE^{m}_d \eta_{mj}(\det(E))^{2/(D-1)}
 \end{equation}

 \begin{multline}
 \label{FirstPoissonBook}
\{P^{ab}(x),  q_{cd}(y)\} = 2|\det(E^c_e)|^{-2/(D-1)} \{( E^a_{k}E^{t}_{m}\eta^{mk}K^{i}_{t}E^b_{i} - E^a_{k}E^{b}_{m}\eta^{mk}K^{i}_{t}E^{t}_{i} ), \; E^j_cE^m_d \eta_{mj} (\det(E))^{2/(D-1)}\} 
 \end{multline}
 All the details of this Poisson bracket calculations can be found in the Appendix B ($\ref{sec:AppendixB}$). Here we just write the result: 
 \begin{equation}
  \{P^{ab}(x),  q_{cd}(y)\} =   2k\delta^b_{(c}\delta^a_{d)}\delta(x, y)
 \end{equation}
 which shows that the symplectic structure is preserved, i.e. the new variables $E^a_{i}$ and $K^{i}_{a}$ are canonical in $SO(2,1)$ case as well. \\[2ex]

 \section{Timelike 3+1 Diffeomorphism and Hamiltonian Constraints}
\label{sec:DiffandHamiltonian}
 Finally we need to check that the hamiltonian and diffeomorphism constraints in new variables commute with the smeared rotational constraint. We introduce the smeared constraint by using $G_{ik}$ form ($\ref{jkconstraint2}$)
\begin{equation}
G(\Lambda) = \int_{\sigma} d^3x \Lambda^{jk}K_{aj}E^a_k,  \;\; \Lambda \in so(2,1)
\end{equation}
Like in $SU(2)$ case the constraints satisfy the Poisson algebra:
\begin{equation}
\{G(\Lambda), G(\Lambda')\} = \frac{k}{2} G([\Lambda, \Lambda'])
\end{equation}
Since the coordinate and momentum ($\ref{pandq}$) are $so(2,1)$ invariant, they will compute with the smeared rotational constraint. \\
Now we need to see if the diffeomorphism and hamiltonian constraints in the new coordinates will commute with the smeared rotational constraint. By substituting the new variables ($\ref{pandq}$) into diffeomorphism and hamiltonian constraints $\cite{Thiemann}$ (1.2.6) and taking s = 1 for $SO(2,1)$ case:
\begin{equation}
H_a = -2q_{ac}D_bP^{bc}
\end{equation}
\begin{equation}
H = - \frac{k}{\sqrt{q}} \left[q_{ac}q_{bd} - \frac{1}{D-1} q_{ab}q_{cd} \right] P^{ab}P^{cd} + \sqrt{q}/k 
\end{equation}
we obtain:
\begin{equation}
\label{Diffeomorphism}
H_a = 2D_b[K^j_a E^b_j - {\delta}^b_a K^j_c E^c_j]
\end{equation}
\begin{equation}
\label{Hamiltonian}
H = -\frac{1}{\sqrt{q}} (K^l_aK^j_b - K^j_aK^l_b)E^a_jE^b_l - \sqrt{q}R
\end{equation}
, where R is a function of $E^a_j$. \\
Both diffeormorphism and hamiltonian constraints commute with the smeared rotational constraint, since both constraints are functions of $q_{ab}$ and $P^{cd}$, and we have showed above that those, when expressed in the new variables, still commute with the smeared rotational constraint. So the whole system of constraints is still first class. 

\section{ Discussion }
\label{sec:Discussion}
We have obtained a new connection in the timelike foliation  with $SO(2,1)$ structure group by reproducing the Sen-Ashtekar-Barbero-Immirzi formalism.  We introduced the new canonically modified variables and proved by  calculating the Poisson brackets that the symplectic structure is preserved. We then introduced the new rotational constraints for $SO(2,1)$ group. By using the isomorphism between the $so(2,1)$ algebra and the algebra $R^3_{2,1}$ with the vector product we were able to obtain a connection, so that the rotational constraint became a Gauss constraint of the new connection with the values in $so(2,1)$ algebra. Interesting that the new connection differs from the $so(3)$ Sen-Ashtekar-Barbero-Immirzi connection not only by $\Gamma^k_a$ and $K^k_a$ taking values in $so(2,1)$ instead of $so(3)$, in which case we would be able to write it in the same way as Sen-Ashtekar-Barbero-Immirzi connection: $A^k_a = \Gamma^k_a + \beta K^k_a$, but also in the component signs due to the presence of the additional coefficient $\eta_{ij}\epsilon_{jkl}$ and the covariant index k in $K_{ak}$ instead of contravariant $K^k_a$ in $so(3)$ case. Thus the $so(2,1)$ connection can be written only in the form: $A^l_{ai} = \eta_{ij}\epsilon_{jkl}(\Gamma^k_a + \beta K_{ka})$. \\[2ex]
Our result has a much simpler form than all previous ones, and clearly shows how $so(2,1)$ connection is different from $so(3)$ one. Also it is the first time that the key-stone fact that makes the whole formalism work in timelike 3+1 case, i.e. $so(2,1) \simeq R^3_{2,1}$ isomorphism and its relation to the $so(2,1)$ connection has been researched. \\[3ex]

\textbf{Acknowledgment}\\
I would like to specially thank Michael Bukatin for the multiple fruitful and challenging discussions. \\[2ex]

\section{Appendix A \quad Momentum-Momentum Poisson Bracket}
\label{sec:AppendixA}

\begin{equation}
P^{ab}(x) = 2|\det(E^c_e)|^{-2/(D-1)}( E^a_kE^{t}_{m}\eta^{mk}K^i_tE^b_i  - E^a_kE^{b}_{m}\eta^{mk}K^i_tE^t_i ) = \frac{2}{q}( E^a_kE^{t}_{m}\eta^{mk}K^i_tE^b_i  - E^a_kE^{b}_{m}\eta^{mk}K^i_tE^t_i )
\end{equation}

\begin{equation}
P^{cd}(x) = 2|\det(E^c_e)|^{-2/(D-1)} ( E^c_kE^{t}_{m}\eta^{mk}K^i_tE^d_i  - E^c_kE^{d}_{m}\eta^{mk}K^i_tE^t_i)= \frac{2}{q}( E^c_kE^{t}_{m}\eta^{mk}K^i_tE^d_i  - E^c_kE^{d}_{m}\eta^{mk}K^i_tE^t_i)
\end{equation}

\begin{multline}
\label{momentumbracket2}
\{P^{ab}(x), P^{cd}(y)\} = \{\frac{2}{q}(E^a_{k_1}E^{t_1}_{p_1}\eta^{p_1k_1}K^{i_1}_{t_1}E^b_{i_1} - E^a_{k_2}E^{b}_{p_2}\eta^{p_2k_2}K^{i_2}_{t_2}E^{t_2}_{i_2}) , \\
 \frac{2}{q}(E^c_{m_1}E^{e_1}_{p_3}\eta^{p_3m_1}K^{j_1}_{e_1}E^d_{j_1} - E^c_{m_2}E^{d}_{p_4}\eta^{p_4m_2}K^{j_2}_{e_2}E^{e_2}_{j_2})
\end{multline}
 By introducing the following notations:\\[2ex]
$a = 1/q$ \\
$b = E^a_{k_1}E^{t_1}_{p_1}\eta^{p_1k_1}K^{i_1}_{t_1}E^b_{i_1}$ \\
$c = E^a_{k_2}E^{b}_{p_2}\eta^{p_2k_2}K^{i_2}_{t_2}E^{t_2}_{i_2}$  \\
$d = 1/q$ \\
$e = E^c_{m_1}E^{e_1}_{p_3}\eta^{p_3m_1}K^{j_1}_{e_1}E^d_{j_1}$ \\
$f = E^c_{m_2}E^{d}_{p_4}\eta^{p_4m_2}K^{j_2}_{e_2}E^{e_2}_{j_2}$\\[2ex]
We can rewrite ($\ref{momentumbracket2}$) as:
\begin{equation}
\label{momentumbracket4}
\{P^{ab}(x), P^{cd}(y)\} =4\{a(b-c), d(e-f)\}
\end{equation}
or by using the Leibniz rule for the Poisson brackets:
\begin{multline}
\label{momentumbracket3}
\{P^{ab}(x), P^{cd}(y)\} =
4( a (\{b, d\} -\{c, d\}) (e-f)+ 
\\
d(\{a, e\} - \{a, f\})(b-c) + 
\\
ad\{b, e\} - ad\{c, e\} - ad\{b, f\} + ad\{c, f\}) + 
\\
\{a, d\}(e-f)(b-c))
\end{multline}
The last term is zero since $\{a, d\} = \{(\det E)^\frac{-2}{D-1}, \{(\det E)^\frac{-2}{D-1}\} = 0$, as $\{E^a_j(x), E^b_k(y)\} = 0$

Let's calculate separately $\{b, f\}$, $\{c, f\}$, $\{c, e\}$, $\{b, e\}$, $\{a, e\}$, $\{a, f\}$, $\{b, d\}$, $\{c, d\}$
\begin{multline}
\label{be0}
\{b, e\} = \{E^a_{k_1}E^{t_1}_{p_1}\eta^{p_1k_1}K^{i_1}_{t_1}E^b_{i_1}, \;  E^c_{m_1} E^{e_1}_{p_3}\eta^{p_3m_1}K^{j_1}_{e_1}E^d_{j_1}\}=
\\
E^a_{k_1}E^{t_1}_{p_1}\eta^{p_1k_1}\{K^{i_1}_{t_1},E^c_{m_1}\}E^b_{i_1}E^{e_1}_{p_3}\eta^{p_3m_1}K^{j_1}_{e_1}E^d_{j_1} + \\
E^a_{k_1}E^{t_1}_{p_1}\eta^{p_1k_1}E^c_{m_1}\{K^{i_1}_{t_1},E^{e_1}_{p_3}\eta^{p_3m_1}\}E^b_{i_1}K^{j_1}_{e_1}E^d_{j_1} + \\
E^a_{k_1}E^{t_1}_{p_1}\eta^{p_1k_1}E^c_{m_1}E^{e_1}_{p_3}\eta^{p_3m_1}K^{j_1}_{e_1}\{K^{i_1}_{t_1},E^d_{j_1}\}E^b_{i_1} +\\
E^c_{m_1}E^{e_1}_{p_3}\eta^{p_3m_1}\{E^a_{k_1},K^{j_1}_{e_1}\}E^d_{j_1}E^{t_1}_{p_1}\eta^{p_1k_1}K^{i_1}_{t_1},E^b_{i_1} + \\
E^c_{m_1}E^{e_1}_{p_3}\eta^{p_3m_1}E^a_{k_1}\{E^{t_1}_{p_1}\eta^{p_1k_1}K^{j_1}_{e_1}\}E^d_{j_1}K^{i_1}_{t_1},E^b_{i_1} +\\
E^c_{m_1}E^{e_1}_{p_3}\eta^{p_3m_1}E^a_{k_1}E^{t_1}_{p_1}\eta^{p_1k_1}K^{i_1}_{t_1}\{E^b_{i_1}, K^{j_1}_{e_1}\}E^d_{j_1}
\end{multline}
\begin{multline}
\label{be}
\{b, e\} =  \{E^a_{k_1}E^{t_1}_{p_1}\eta^{p_1k_1}K^{i_1}_{t_1}E^b_{i_1}, \;  E^c_{m_1} E^{e_1}_{p_3}\eta^{p_3m_1}K^{j_1}_{e_1}E^d_{j_1}\}=
\\
E^a_{k_1}E^{t_1}_{p_1}\eta^{p_1k_1}(-\frac{k}{2}\delta^{i_1}_{m_1}\delta^c_{t_1})E^b_{i_1}E^{e_1}_{p_3}\eta^{p_3m_1}K^{j_1}_{e_1}E^d_{j_1} + \\
E^a_{k_1}E^{t_1}_{p_1}\eta^{p_1k_1}E^c_{m_1}(-\frac{k}{2}\delta^{i_1}_{m_1}\delta^{e_1}_{p_3}\eta^{p_3t_1})E^b_{i_1}K^{j_1}_{e_1}E^d_{j_1} + \\
E^a_{k_1}E^{t_1}_{p_1}\eta^{p_1k_1}E^c_{m_1}E^{e_1}_{p_3}\eta^{p_3m_1}K^{j_1}_{e_1}(-\frac{k}{2}\delta^{i_1}_{j_1}\delta^d_{t_1})E^b_{i_1} +\\
E^c_{m_1}E^{e_1}_{p_3}\eta^{p_3m_1}(\frac{k}{2}\delta^{j_1}_{k_1}\delta^a_{e_1})E^d_{j_1}E^{t_1}_{p_1}\eta^{p_1k_1}K^{i_1}_{t_1}E^b_{i_1} + \\
E^c_{m_1}E^{e_1}_{p_3}\eta^{p_3m_1}E^a_{k_1}(\frac{k}{2}\delta^{t_1}_{e_1}\delta^{j_1}_{p_1})\eta^{p_1k_1}E^d_{j_1}K^{i_1}_{t_1}E^b_{i_1} +\\
E^c_{m_1}E^{e_1}_{p_3}\eta^{p_3m_1}E^a_{k_1}E^{t_1}_{p_1}\eta^{p_1k_1}K^{i_1}_{t_1}(\frac{k}{2}\delta^{j_1}_{i_1}\delta^b_{e_1})E^d_{j_1} = 
\\
q^2\frac{k}{2}(-q^{ac}q^{be_1}K^{j_1}_{e_1}E^d_{j_1} - q^{ae_1}q^{bc}K^{j_1}_{e_1}E^d_{j_1} - q^{ad}q^{ce_1}K^{j_1}_{e_1}E^b_{j_1} + q^{ca}q^{dt_1}K^{i_1}_{t_1}E^b_{i_1} 
\\
+q^{ct_1}q^{da}K^{i_1}_{t_1}E^b_{i_1} + q^{cb}q^{at_1}K^{i_1}_{t_1}E^d_{i_1}) =
\\
=q^2\frac{k}{2}q( -q^{ac} \hat{G}^{bd} - q^{bc} \hat{G}^{ad} - q^{ad} \hat{G}^{cb} + q^{ac} \hat{G}^{db} + q^{da} \hat{G}^{cb} + q^{cb} \hat{G}^{ad}) = 
\\ = \frac{k}{2}q^3q^{ac}( \hat{G}^{db} - \hat{G}^{bd} ) = \frac{k}{2}q^3q^{ac}G^{db}
\end{multline}
,where
\begin{equation}
G^{db} := \hat{G}^{db} - \hat{G}^{bd} 
\end{equation}
and we have introduced the notations for $\hat{G}$ with various indices:
\begin{equation}
 \hat{G}^{db} = q^{de_1}K^{j_1}_{e_1}q^{bp}E^{j_1}_{p}  
\end{equation}
 Thus
\begin{equation}
\label{bfResult}
\{b, e\} =  \frac{k}{2}q^3q^{ac}G^{db} 
\end{equation}
The next Poisson bracket is:
\begin{multline}
\label{ae2}
\{a, e\} = \{(\det E)^\frac{-2}{D-1}, \;   E^c_{m_1}E^{e_1}_{p_3}\eta^{p_3m_1}K^{j_1}_{e_1}E^d_{j_1}\}=
\\
E^c_{m_1}E^{e_1}_{p_3}\eta^{p_3m_1}\{ (\det E)^\frac{-2}{D-1}, K^{j_1}_{e_1}\}E^d_{j_1} = 
\\
E^c_{m_1}E^{e_1}_{p_3}\eta^{p_3m_1}\frac{-2}{D-1}  (\det E)^\frac{-2}{D-1} \frac{\{ (\det E), K^{j_1}_{e_1}\}}{\det E}E^d_{j_1}=
\\
E^c_{m_1}E^{e_1}_{p_3}\eta^{p_3m_1}\frac{-2}{D-1} \frac{1}{q} E^n_r \{E^r_n, K^{j_1}_{e_1} \}E^d_{j_1} = 
\\
q^{ce_1} q \frac{-2}{D-1} \frac{1}{q}E^n_r (\frac{k}{2}\delta^{j_1}_n \delta_{e_1}^r) E^d_{j_1}=
\\
q^{ce_1} E^d_{j_1}\frac{-k}{D-1} E^{j_1}_{e_1} = q^{ce_1} \delta^d_{e_1} \frac{-k}{D-1} =  q^{cd}\frac{-k}{D-1}
\end{multline}
We obtain:
\begin{equation}
\label{aeResult1}
\{a, e\} =  q^{cd}\frac{-k}{D-1}
\end{equation}
The next bracket can be obtained from ($\ref{ae2}$), by changing the sign and making the following index replacement:
\begin{equation}
c \rightarrow a, \; d \rightarrow b, \; a \rightarrow c, \; b \rightarrow d
\end{equation}
\begin{equation}
\label{bd2}
\{b, d\} = \{E^a_{k_1}E^{t_1}_{p_1}\eta^{p_1k_1}K^{i_1}_{t_1}E^b_{i_1}, \; (\det E)^\frac{-2}{D-1}\}
\end{equation}
\begin{equation}
\label{bdResult2}
\{b, d\} =  q^{ab}\frac{k}{D-1}
\end{equation}
The next bracket goes as follows:
\begin{multline}
\{b, f\} = \{ E^a_{k_1}E^{t_1}_{p_1}\eta^{p_1k_1}K^{i_1}_{t_1}E^b_{i_1}, \; E^c_{m_2}E^{d}_{p_4}\eta^{p_4m_2}K^{j_2}_{e_2}E^{e_2}_{j_2}\}=
\\
E^a_{k_1}E^{t_1}_{p_1}\eta^{p_1k_1}\{K^{i_1}_{t_1},E^{c}_{m_2}\}E^b_{i_1}E^{d}_{p_4}\eta^{p_4m_2}K^{j_2}_{e_2}E^{e_2}_{j_2} + 
\\
E^a_{k_1}E^{t_1}_{p_1}\eta^{p_1k_1}E^c_{m_2}\{K^{i_1}_{t_1},E^{d}_{p_4}\eta^{p_4m_2}\}E^b_{i_1}K^{j_2}_{e_2}E^{e_2}_{j_2} + 
\\
E^a_{k_1}E^{t_1}_{p_1}\eta^{p_1k_1}E^c_{m_2}E^{d}_{p_4}\eta^{p_4m_2}K^{j_2}_{e_2}\{K^{i_1}_{t_1},E^{e_2}_{j_2}\}E^b_{i_1} + 
\\
E^c_{m_2}E^{d}_{p_4}\eta^{p_4m_2}\{E^{a}_{k_1}, K^{j_2}_{e_2}\}E^{e_2}_{j_2}E^{t_1}_{p_1}\eta^{p_1k_1}K^{i_1}_{t_1}E^{b}_{i_1} +
\\
E^c_{m_2}E^{d}_{p_4}\eta^{p_4m_2}E^a_{k_1}\{E^{t_1}_{p_1}\eta^{p_1k_1}, K^{j_2}_{e_2}\}E^{e_2}_{j_2}K^{i_1}_{t_1}E^{b}_{i_1} +
\\
E^c_{m_2}E^{d}_{p_4}\eta^{p_4m_2}E^a_{k_1}E^{t_1}_{p_1}\eta^{p_1k_1}K^{i_1}_{t_1}\{E^{b}_{i_1}, K^{j_2}_{e_2}\}E^{e_2}_{j_2}
\end{multline}
or
\begin{multline}
\label{bf4}
\{b, f\} = \{ E^a_{k_1}E^{t_1}_{p_1}\eta^{p_1k_1}K^{i_1}_{t_1}E^b_{i_1}, \; E^c_{m_2}E^{d}_{p_4}\eta^{p_4m_2}K^{j_2}_{e_2}E^{e_2}_{j_2}\} =
\\
E^a_{k_1}E^{t_1}_{p_1}\eta^{p_1k_1}(-\frac{k}{2}\delta^{i_1}_{m_2} \delta^{c}_{t_1} )E^b_{i_1}E^{d}_{p_4}\eta^{p_4m_2}K^{j_2}_{e_2}E^{e_2}_{j_2} + 
\\
E^a_{k_1}E^{t_1}_{p_1}\eta^{p_1k_1}E^c_{m_2}(-\frac{k}{2}\delta^{i_1}_{p_4} \delta^{d}_{t_1} )\eta^{p_4m_2}E^b_{i_1}K^{j_2}_{e_2}E^{e_2}_{j_2} + 
\\
E^a_{k_1}E^{t_1}_{p_1}\eta^{p_1k_1}E^c_{m_2}E^{d}_{p_4}\eta^{p_4m_2}K^{j_2}_{e_2}(-\frac{k}{2}\delta ^{i_1}_{j_2} \delta^{e_2}_{t_1})E^b_{i_1} + 
\\
E^c_{m_2}E^{d}_{p_4}\eta^{p_4m_2}(\frac{k}{2}\delta^{a}_{e_2} \delta^{j_2}_{k_1})E^{e_2}_{j_2}E^{t_1}_{p_1}\eta^{p_1k_1}K^{i_1}_{t_1}E^{b}_{i_1} +
\\
E^c_{m_2}E^{d}_{p_4}\eta^{p_4m_2}E^a_{k_1}(\frac{k}{2}\delta^{t_1}_{e_2} \delta^{j_2}_{p_1} )\eta^{p_1k_1}E^{e_2}_{j_2}K^{i_1}_{t_1}E^{b}_{i_1} +
\\
E^c_{m_2}E^{d}_{p_4}\eta^{p_4m_2}E^a_{k_1}E^{t_1}_{p_1}\eta^{p_1k_1}K^{i_1}_{t_1}(\frac{k}{2}\delta^{b}_{e_2} \delta^{j_2}_{i_1}\}E^{e_2}_{j_2} =
\\
\frac{k}{2}(- q^{ac}qq^{bd}qK^{j_2}_{e_2}E^{e_2}_{j_2} - q q^{ad}qq^{bc}K^{j_2}_{e_2}E^{e_2}_{j_2} - q^{ae_2}qq^{cd}qK^{j_2}_{e_2}E^{b}_{j_2} +
\\
q^{cd}qq^{at_1}qK^{i_1}_{t_1}E^b_{i_1} + q^{cd}qq^{at_1}qK^{i_1}_{t_1}E^b_{i_1} + 
q^{cd}qq^{at_1}qK^{i_1}_{t_1}E^b_{i_1} ) = 
\\
\frac{-k q^2}{2}K^{j_2}_{e_2}E^{e_2}_{j_2}(q^{ac}q^{bd} + q^{ad}q^{bc}) + kq^2q^{cd}q^{at_1}K^{i_1}_{t_1}E^b_{i_1}
\end{multline}
To summarize 
\begin{multline}
\label{bf4Result}
\{b, f\} =\{ E^a_{k_1}E^{t_1}_{p_1}\eta^{p_1k_1}K^{i_1}_{t_1}E^b_{i_1}, \; E^c_{m_2}E^{d}_{p_4}\eta^{p_4m_2}K^{j_2}_{e_2}E^{e_2}_{j_2}\}
\\
=-\frac{kq^2}{2}K^{j_2}_{e_2}E^{e_2}_{j_2}(q^{ac}q^{bd} + q^{ad}q^{bc}) + kq^2q^{cd}q^{at_1}K^{i_1}_{t_1}E^b_{i_1}
\\
= -\frac{kq^2}{2}K^{j_2}_{e_2}E^{e_2}_{j_2}(q^{ac}q^{bd} + q^{ad}q^{bc}) + kq^2q^{cd}q^{at_1}K^{i_1}_{t_1}qq^{bp}E^{i_1}_p
\\
=  -\frac{kq^2}{2}K^{j_2}_{e_2}E^{e_2}_{j_2}(q^{ac}q^{bd} + q^{ad}q^{bc}) + kq^3q^{cd}K^{ai_1}E^{bi_1}
\end{multline}
The bracket $\{c, e\}$ is similar to $\{f, b\} = -\{b, f\}$ above in ($\ref{bf4}$) with the following index replacement:
\begin{equation}
a \rightarrow c, \; b \rightarrow d, \; c \rightarrow a, \; d \rightarrow b
\end{equation}
\begin{multline}
\label{ce}
\{c, e\} = \{E^a_{k_2}E^{b}_{p_2}\eta^{p_2k_2}K^{i_2}_{t_2}E^{t_2}_{i_2},  \; E^c_{m_1}E^{e_1}_{p_3}\eta^{p_3m_1}K^{j_1}_{e_1}E^d_{j_1}\} = -\{E^c_{m_1}E^{e_1}_{p_3}\eta^{p_3m_1}K^{j_1}_{e_1}E^d_{j_1}, \; E^a_{k_2}E^{b}_{p_2}\eta^{p_2k_2}K^{i_2}_{t_2}E^{t_2}_{i_2}\}
\\
=\frac{kq^2}{2}K^{j_2}_{e_2}E^{e_2}_{j_2}(q^{ca}q^{db} + q^{da}q^{cb}) - kq^2q^{ab}q^{ct_1}K^{i_1}_{t_1}E^d_{i_1}
\\
=\frac{kq^2}{2}K^{j_2}_{e_2}E^{e_2}_{j_2}(q^{ca}q^{db} + q^{cb}q^{da}) - kq^3q^{ab}K^{ci_1}E^{di_1}
\end{multline}

\begin{multline}
\label{af6}
\{a, f\} = \{ {(\det E)}^{\frac{-2}{D-1}}, \; E^c_{m_2}E^{d}_{p_4}\eta^{p_4m_2}K^{j_2}_{e_2}E^{e_2}_{j_2}\} = 
E^c_{m_2}E^{d}_{p_4}\eta^{p_4m_2}\frac{-2}{D-1} \frac{1}{q} \frac{\{ (\det E), K^{j_2}_{e_2}\} }{(\det E)} E^{e_2}_{j_2}= 
\\
E^c_{m_2}E^{d}_{p_4}\eta^{p_4m_2}\frac{-2}{D-1} \frac{1}{q} E^{j_3}_{e_3}\{E^{e_3}_{j_3}, K^{j_2}_{e_2}\}E^{e_2}_{j_2} = 
E^c_{m_2}E^{d}_{p_4}\eta^{p_4m_2}\frac{-2}{D-1} \frac{1}{q} E^{j_3}_{e_3}\frac{k}{2}\delta^{j_2}_{j_3} \delta^{e_2}_{e_3}E^{e_2}_{j_2} =
\\
\frac{-k}{D-1} q q^{cd} \frac{1}{q} E^{j_2}_{e_2}E^{e_2}_{j_2} = \frac{-k}{D-1} q^{cd}D
\end{multline}
The bracket $\{c, d\}$ can be calculated from ($\ref{af6}$) by changing the sign and making the following index replacement:
\begin{equation}
c \rightarrow a, \; d \rightarrow b, \; a \rightarrow c, \; b \rightarrow d
\end{equation}
We  obtain:
\begin{equation}
\label{cd6}
\{c, d\} = \{E^a_{k_2}E^{b}_{p_2}\eta^{p_2k_2}K^{i_2}_{t_2}E^{t_2}_{i_2}, \; (\det E)^{\frac{-2}{D-1}}\} = \frac{k}{D-1}  q^{ab}D
\end{equation}
Finally we need to calculate the last bracket $\{c, f\}$:
\begin{multline}
\{c, f\} = \{E^a_{k_2}E^{b}_{p_2}\eta^{p_2k_2}K^{i_2}_{t_2}E^{t_2}_{i_2}, \; E^c_{m_2}E^{d}_{p_4}\eta^{p_4m_2}K^{j_2}_{e_2}E^{e_2}_{j_2}\}=
\\
E^a_{k_2}E^{b}_{p_2}\eta^{p_2k_2}\{K^{i_2}_{t_2}, E^c_{m_2}\}E^{t_2}_{i_2}E^{d}_{p_4}\eta^{p_4m_2}K^{j_2}_{e_2}E^{e_2}_{j_2} + 
\\
E^a_{k_2}E^{b}_{p_2}\eta^{p_2k_2}E^c_{m_2}\{K^{i_2}_{t_2}, E^{d}_{p_4}\eta^{p_4m_2}\}E^{t_2}_{i_2}K^{j_2}_{e_2}E^{e_2}_{j_2} + 
\\
E^a_{k_2}E^{b}_{p_2}\eta^{p_2k_2}E^c_{m_2}E^{d}_{p_4}\eta^{p_4m_2}K^{j_2}_{e_2}\{K^{i_2}_{t_2}, E^{e_2}_{j_2}\}E^{t_2}_{i_2} +
\\
E^c_{m_2}E^{d}_{p_4}\eta^{p_4m_2}\{E^a_{k_2}, K^{j_2}_{e_2}\}E^{e_2}_{j_2}E^{b}_{p_2}\eta^{p_2k_2}K^{i_2}_{t_2}E^{t_2}_{i_2} + 
\\
E^c_{m_2}E^{d}_{p_4}\eta^{p_4m_2}E^a_{k_2}\{E^{b}_{p_2}\eta^{p_2k_2}, K^{j_2}_{e_2}\}E^{e_2}_{j_2}K^{i_2}_{t_2}E^{t_2}_{i_2} + 
\\
E^c_{m_2}E^{d}_{p_4}\eta^{p_4m_2}E^a_{k_2}E^{b}_{p_2}\eta^{p_2k_2}K^{i_2}_{t_2}\{E^{t_2}_{i_2}, K^{j_2}_{e_2}\}E^{e_2}_{j_2}
\end{multline}
or
\begin{multline}
\{c, f\} = \{E^a_{k_2}E^{b}_{p_2}\eta^{p_2k_2}K^{i_2}_{t_2}E^{t_2}_{i_2}, \; E^c_{m_2}E^{d}_{p_4}\eta^{p_4m_2}K^{j_2}_{e_2}E^{e_2}_{j_2}\} =
\\
E^a_{k_2}E^{b}_{p_2}\eta^{p_2k_2}E^{t_2}_{i_2}(-\frac{k}{2}\delta^{i_2}_{m_2}\delta^c_{t_2})E^{d}_{p_4}\eta^{p_4m_2}K^{j_2}_{e_2}E^{e_2}_{j_2} + 
\\
E^a_{k_2}E^{b}_{p_2}\eta^{p_2k_2}E^{t_2}_{i_2}E^c_{m_2}(- \frac{k}{2}\delta^{i_2}_{p_4} \delta^d_{t_2})\eta^{p_4m_2}K^{j_2}_{e_2}E^{e_2}_{j_2} + 
\\
E^a_{k_2}E^{b}_{p_2}\eta^{p_2k_2}E^{t_2}_{i_2}E^c_{m_2}E^{d}_{p_4}\eta^{p_4m_2}K^{j_2}_{e_2}(- \frac{k}{2} \delta^{i_2}_{j_2}\delta^{e_2}_{t_2}) +
\\
E^c_{m_2}E^{d}_{p_4}\eta^{p_4m_2}( \frac{k}{2} \delta^{j_2}_{k_2}\delta^a_{e_2})E^{e_2}_{j_2}E^{b}_{p_2}\eta^{p_2k_2}K^{i_2}_{t_2}E^{t_2}_{i_2} + 
\\
E^c_{m_2}E^{d}_{p_4}\eta^{p_4m_2}E^a_{k_2}( \frac{k}{2} \delta^{j_2}_{p_2}\delta^b_{e_2})E^{e_2}_{j_2}\eta^{p_2k_2}K^{i_2}_{t_2}E^{t_2}_{i_2} + 
\\
E^c_{m_2}E^{d}_{p_4}\eta^{p_4m_2}E^a_{k_2}E^{b}_{p_2}\eta^{p_2k_2}K^{i_2}_{t_2}(\frac{k}{2} \delta^{t_2}_{e_2}\delta^{j_2}_{i_2})E^{e_2}_{j_2} = 
\\
\frac{k}{2}(-q^{ab}qq^{cd}qK^{j_2}_{e_2}E^{e_2}_{j_2} - q^{ab}qq^{cd}qK^{j_2}_{e_2}E^{e_2}_{j_2} - q^{ab}qq^{cd}qK^{j_2}_{e_2}E^{e_2}_{j_2} 
\\
+ q^{cd}qq^{ab}qK^{i_2}_{t_2}E^{t_2}_{i_2} +  q^{cd}qq^{ab}qK^{i_2}_{t_2}E^{t_2}_{i_2} +  q^{cd}qq^{ab}qK^{i_2}_{t_2}E^{t_2}_{i_2}) = 0
\end{multline}
Thus
\begin{equation}
\label{cf}
\{c, f\} = \{E^a_{k_2}E^{b}_{p_2}\eta^{p_2k_2}K^{i_2}_{t_2}E^{t_2}_{i_2}, \; E^c_{m_2}E^{d}_{p_4}\eta^{p_4m_2}K^{j_2}_{e_2}E^{e_2}_{j_2}\}=0
\end{equation}
To summarize we have obtained:\\[2ex]
$\{b, d\} =  \frac{k}{D-1}q^{ab}$\\
$\{c, d\} = \frac{k}{D-1} q^{ab}D$\\
$\{a, e\} =  \frac{-k}{D-1}q^{cd}$\\
$\{a, f\} = \frac{-k}{D-1} q^{cd}D$\\
$\{b, e\} = \frac{k}{2}qq^{ac}G^{db}$\\
$\{c, e\} = \frac{kq^2}{2}K^{j_2}_{e_2}E^{e_2}_{j_2}(q^{ca}q^{db} + q^{da}q^{cb}) - kq^3q^{ab}K^{ci_1}E^{di_1}$\\
$\{b, f\} = -\frac{kq^2}{2}K^{j_2}_{e_2}E^{e_2}_{j_2}(q^{ac}q^{bd} + q^{ad}q^{bc}) + kq^3q^{cd}K^{ai_1}E^{bi_1}$\\
$\{c, f\} = 0$\\[2ex]

By substituting into ($\ref{momentumbracket3}$) :
\begin{multline}
\label{momentumbracket7}
\{P^{ab}(x), P^{cd}(y)\} = 4( a (\{b, d\} - \{c, d\})(e-f) + 
\\
d(\{a, e\} -\{a, f\})(b-c) + 
\\
ad\{b, e\} - ad\{c, e\} - ad\{b, f\} + ad\{c, f\})
\end{multline}
we obtain:
\begin{multline}
 \{P^{ab}(x), P^{cd}(y)\} = \frac{4}{q}(\frac{k}{D-1}q^{ab} - \frac{k}{D-1} q^{ab}D)(E^c_{m_1}E^{e_1}_{p_3}\eta^{p_3m_1}K^{j_1}_{e_1}E^d_{j_1} - E^c_{m_2}E^{d}_{p_4}\eta^{p_4m_2}K^{j_2}_{e_2}E^{e_2}_{j_2})
  + 
\\
\frac{4}{q} 
 (\frac{-k}{D-1}q^{cd} -  \frac{-k}{D-1} q^{cd}D) (E^a_{k_1}E^{t_1}_{p_1}\eta^{p_1k_1}K^{i_1}_{t_1}E^b_{i_1} - E^a_{k_2}E^{b}_{p_2}\eta^{p_2k_2}K^{i_2}_{t_2}E^{t_2}_{i_2}) 
 \\
 +\frac{k}{2}\frac{4}{q^2}  q^3q^{ac}G^{db} 
 \\-\frac{4k}{2q^2} (q^2K^{j_2}_{e_2}E^{e_2}_{j_2}(q^{ca}q^{db} + q^{da}q^{cb}) - 2q^2q^{ab}q^{ct_1}K^{i_1}_{t_1}E^d_{i_1}) 
 \\
 -\frac{4k}{2q^2} (-q^2K^{j_2}_{e_2}E^{e_2}_{j_2}(q^{ac}q^{bd} + q^{ad}q^{bc}) + 2q^2q^{cd}q^{at_1}K^{i_1}_{t_1}E^b_{i_1}))
 \end{multline}
\begin{multline}
 \{P^{ab}(x), P^{cd}(y)\} =  \frac{4}{q}(qq^{ce_1}K^{j_1}_{e_1}E^d_{j_1} - q q^{cd}K^{j_2}_{e_2}E^{e_2}_{j_2})(-k q^{ab})
+  \frac{4}{q}(qq^{at_1}K^{i_1}_{t_1}E^b_{i_1} - qq^{ab}K^{i_2}_{t_2}E^{t_2}_{i_2}))(k q^{cd} ) 
\\
+2kqq^{ac}G^{db} 
\\
+
\frac{4k}{q^2}(q^2q^{ab}K^{ci_1}E^{d}_{i_1} -  q^2q^{cd}K^{ai_1}E^{b}_{i_1} )
\end{multline}
\begin{multline}
 \{P^{ab}(x), P^{cd}(y)\} =  -4k q^{ab}K^{cj_1}E^d_{j_1}
+  4k q^{cd} K^{ai_1}E^b_{i_1} 
+ 2kq q^{ac}G^{db} 
\\
+
4k(q^{ab}K^{ci_1}E^{d}_{i_1} -  q^{cd}K^{ai_1}E^{b}_{i_1} ) = 2kqq^{ac}G^{bd}
\end{multline}

\begin{equation}
 \{P^{ab}(x), P^{cd}(y)\} = 2kqq^{ac}G^{bd}
\end{equation}

\section{Appendix B \quad Coordinate-Momentum Poisson Bracket }
\label{sec:AppendixB}

We mark each line by the label L(line number) and provide the detailed comments underneath the formula on how we move from one line to the next in our calculations.
\begin{equation}
P^{ab}(x) = 2|\det(E^c_e)|^{-2/(D-1)}( E^a_{k_1}E^{t_1}_{m_1}\eta^{m_1k_1}K^{i_1}_{t_1}E^b_{i_1} - E^a_{k_2}E^{b}_{m_2}\eta^{m_2k_2}K^{i_2}_{t_2}E^{t_2}_{i_2} ) 
\end{equation}
 \begin{equation}
 q_{cd}(y) = E^{i_3}_cE^{j_3}_d \eta_{i_3j_3}(\det(E))^{2/(D-1)}
 \end{equation}
 \begin{multline}
 \label{FirstPoissonBook}
 \mbox{L1:  }\, \{P^{ab}(x),  q_{cd}(y)\} = 2|\det(E^c_e)|^{-2/(D-1)} \{( E^a_{k_1}E^{t_1}_{m_1}\eta^{m_1k_1}K^{i_1}_{t_1}E^b_{i_1} - E^a_{k_2}E^{b}_{m_2}\eta^{m_2k_2}K^{i_2}_{t_2}E^{t_2}_{i_2} ) , \; 
 \\ E^{i_3}_cE^{j_3}_d \eta_{i_3j_3}(\det(E))^{2/(D-1)}\} 
 \\
    \mbox{L2:  }\,=\frac{2}{q} \{ (E^a_{k_1}E^t_{m_1} \eta^{m_1k_1}  K^{i_1}_{t_1} qq^{be}E^{j_1}_e\eta_{i_1j_1}  - E^a_{k_2}E^b_{m_2} \eta^{m_2k_2} K^{i_2}_{t_2} qq^{t_2e}E^{j_2}_e\eta_{i_2j_2}),  \; E^{i_3}_cE^{j_3}_d\eta_{i_3j_3} (\det(E))^{2/(D-1)}\}
  \\
 =\frac{2}{q}q^2(q^{at}q^{be} - q^{ab}q^{te})E^j_e\eta_{ij}\{K^i_t, \; E^{i_3}_cE^{j_3}_d \eta_{i_3j_3} (\det(E))^{2/(D-1)}\}
\\
   \mbox{L3:  }\, =2q(q^{at}q^{be} - q^{ab}q^{te})E^j_e\eta_{ij}(q\{K^i_tE^{i_3}_c\}E^{j_3}_d\eta_{i_3j_3} + q\{K^i_tE^{j_3}_d\}E^{i_3}_c\eta_{i_3j_3} + \frac{2}{D-1} \frac{q_{cd}}{q}q \frac{\{K^i_t, \det(E)\}}{\det(E)}) =
 \\
  \mbox{L4:  }\, 2q(q^{at}q^{be} - q^{ab}q^{te})E^j_e\eta_{ij}(q(-E^j_c)\{K^i_t, E^p_j\}E^{i_3}_pE^{j_3}_d\eta_{i_3j_3} + q(-E^j_d)\{K^i_t, E^p_j\}E^{j_3}_pE^{i_3}_c\eta_{i_3j_3}+
 \\
 \frac{2}{D-1} q_{cd} \{K^i_t, E^m_j\}E^j_m) = 
 \\
   \mbox{L5:  }\, 2q(q^{at}q^{be} - q^{ab}q^{te})E^j_e\eta_{ij}(q(-E^j_c)(-\frac{k}{2} \delta^i_j \delta^p_t) E^{i_3}_pE^{j_3}_d\eta_{i_3j_3} + q(-E^j_d)(-\frac{k}{2} \delta^i_j \delta^p_t) E^{j_3}_pE^{i_3}_c\eta_{i_3j_3}+
 \\
 \frac{2}{D-1} q_{cd} (-\frac{k}{2} \delta^i_j \delta^m_t ) E^j_m) = 
 \\
  \mbox{L6:  }\, kq(q^{at}q^{be} - q^{ab}q^{te})E^j_e\eta_{ij}(qE^i_cE^{i_3}_tE^{j_3}_d\eta_{i_3j_3} + qE^i_dE^{j_3}_tE^{i_3}_c\eta_{i_3j_3} - \frac{2}{D-1} q_{cd} E^i_t) =
  \\
   \mbox{L7:  }\,  kq(q^{at}q^{be} - q^{ab}q^{te})(q \frac{q_{ec}}{q} \frac{ q_{dt}}{q} + q \frac{q_{ed}}{q} \frac{q_{ct}}{q} - 
   \frac{2}{D-1} q_{cd} \frac{q_{et}}{q}) =
   \\
   \mbox{L8:  }\,  k (q^{at}q^{be} - q^{ab}q^{te})(q_{ec}q_{dt} + q_{ed}q_{ct} - \frac{2}{D-1} q_{cd}q_{et})=
   \\
    \mbox{L9:  }\, k(\delta^b_c \delta^a_d + \delta^b_d \delta^a_c) - 2q^{ab}q_{cd}q - \frac{2}{D-1}(q_{cd}q^{ab} - D q_{cd}q^{ab})=
   \\
 \mbox{L10:  }\, k(\delta^b_c \delta^a_d + \delta^b_d \delta^a_c) - 2q^{ab}q_{cd}q - \frac{2}{D-1}(1 - D )q_{cd}q^{ab} = 
   \\
  \mbox{L11:  }\, k(\delta^b_c \delta^a_d + \delta^b_d \delta^a_c) - 2q^{ab}q_{cd}q + 2q^{ab}q_{cd}q = 
   \\
   \mbox{L12:  }\,  k2\delta^b_{(c}\delta^a_{d)}\delta(x, y)
 \end{multline}
 \begin{equation}
  \{P^{ab}(x),  q_{cd}(y)\} =   2k\delta^b_{(c}\delta^a_{d)}\delta(x, y)
 \end{equation}
 ,where\\
 in the line $\mbox{L2:}$ we used  $E^a_i E^b_j \eta^{ij} = q q^{ab}$ and $ q := (\det(E))^{2/(D-1)}$\\
 in the line  $\mbox{L3:}$ we used Leibniz rule and $\; \mbox{and} \{E^a_j(x), E^b_k(y)\} = 0$\\
 in the line $\mbox{L4:}$ we used: $\delta E^i_a = -E^i_a \delta E^b_i E^i_b$ and $ [\delta(E)]/\det(E) = E^j_a\delta E^a_j$\\
 in the line $\mbox{L5:}$ we calculated the Poisson brackets: $ \{E^a_i(x), K^j_b(y)\} = \frac{k}{2}\delta^a_b\delta^j_i\delta(x, y)$\\
 in the line $\mbox{L7:}$ we used: $E^i_e E^j_c \eta_{ij}= q_{ec}/q $, etc, \\
 in the line $\mbox{L9}$: we have opened the parenthesis and used:
 $q^{at} q_{td} = \delta^a_d$ and  $q^{et} q_{et} = D$\\
in the  line $\mbox{L10:}$ $D-1$ cancels.\\
in the  line $\mbox{L11:}$ the last two terms are the same and mutually cancel.

\end{document}